\documentclass{sig-alternate-05-2015}

\usepackage{hyperref}
\usepackage{color}

\newcommand{\ie}{{\emph{i.e.}}}
\newcommand{\eg}{{\emph{e.g.}}}
\newcommand{\mypara}[1]{\vspace{0.06in}\noindent\textit{\textbf{{#1}.}}}

\newcommand{\sns}{Sketch-n-Sketch}
\newcommand{\powerpoint}{{PowerPoint}}

\begin{document}
\CopyrightYear{2016} 
\setcopyright{acmlicensed}
\conferenceinfo{ICSE '16 Companion,}{May 14--22, 2016, Austin, TX, USA}
\isbn{978-1-4503-4205-6/16/05}\acmPrice{\$15.00}
\doi{http://dx.doi.org/10.1145/2889160.2889210}

\title{Prodirect Manipulation:\\
       Bidirectional Programming for the Masses}
\numberofauthors{1} 
\author{
\alignauthor
Ravi Chugh\\
       \affaddr{Department of Computer Science, University of Chicago}\\
       \affaddr{Ryerson Laboratory, 1100 E 58th Street, Chicago, IL, USA 60637}\\
       \email{rchugh@cs.uchicago.edu}
}

\maketitle
\begin{abstract}
Software interfaces today generally fall at either end of
a spectrum. On one end are programmable systems, which allow expert
users (\ie{} programmers) to write software artifacts that describe
complex abstractions, but programs are disconnected from their eventual
output. On the other end are domain-specific graphical user
interfaces (GUIs), which allow end users (\ie{} non-programmers)
to easily create varied content but present
insurmountable walls when a desired feature is not built-in.
Both programmatic and direct manipulation have distinct
strengths, but users must typically choose one over the other or
use some ad-hoc combination of systems.
Our goal, put simply, is to bridge this divide.

We envision novel software systems that tightly couple
programmatic and direct manipulation --- a combination we dub
\emph{prodirect manipulation} --- for a variety of use cases.
This will require advances in a broad range of software
engineering disciplines, from program analysis and program synthesis
technology
to user interface design and evaluation. In this extended
abstract, we propose two general strategies
--- \emph{real-time program synthesis}
and \emph{domain-specific synthesis of general-purpose programs} ---
that may prove fruitful for overcoming the technical challenges.
We also discuss metrics that will be important in
evaluating the usability and utility of prodirect manipulation systems.
\end{abstract}

\category{D.3.3}{Programming Languages}{Language Constructs and Features}
\category{H.5.2}{Information Systems Applications}{User Interfaces}
\category{D.2.6}{Software Engineering}{Programming Environments}
\category{F.3.2}{Logics and Meanings of Programs}{Program Analysis}

\keywords{Prodirect Manipulation, Program Synthesis, \\
          End User Programming, Human-Computer Interaction, \\
          Bidirectional Programming}

\section{INTRODUCTION} \noindent
\emph{Direct manipulation} describes interfaces that incorporate
``visibility of the object of interest; rapid, reversible, incremental actions;
 and replacement of complex command language syntax by
direct manipulation of the object of interest''~\cite{DM}.
Such interfaces are developed for a variety of domains in which objects
have inherently visual representations, including word processors
(\eg{} Microsoft Word, Apple Pages, and Google Docs),
presentation systems
(\eg{} Microsoft PowerPoint, Apple Keynote, and Google Slides),
vector graphics editors (\eg{} Adobe Illustrator), and
user interface (UI) design tools
(\eg{} Adobe Dreamweaver and Apple XCode Interface Builder).
After prototyping phases, however, relying solely on direct
manipulation can lead to repetitive copy-and-paste tasks
and, furthermore, can make it difficult for expert users to manipulate
complex content in reusable and composable ways.

At the other end of the spectrum are purely programmatic systems
(\eg{} \LaTeX{} for document layout,
Slideshow~\cite{SlideshowICFP04} for presentations,
Processing~(\url{processing.org}) for visual arts, and
Apple Swift for UI design).
In these systems, users define their content with high-level,
general-purpose programs, which have access to the powerful
abstraction capabilities that are afforded by programming.
At the same time, however, it can be difficult and unintuitive
to perform stylistic changes.
As a result, programmers often enter a
tedious cycle of changing parameters in the program, running it again,
inspecting the newly rendered output, and repeating until the desired
change has been affected.

\begin{figure}[t]
\centering
\includegraphics[scale=0.35]{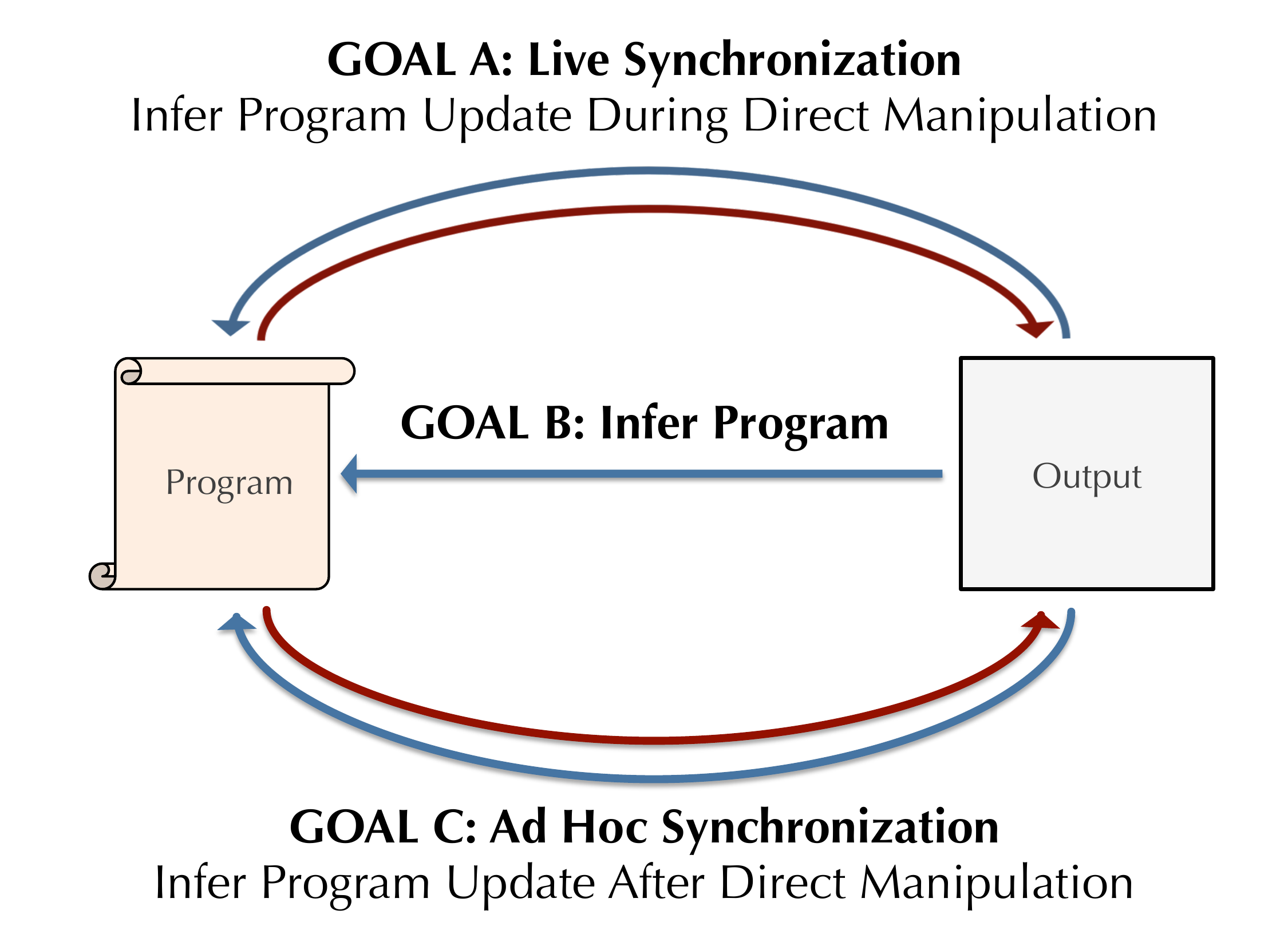}
\caption{Three Desired Modes of Use}
\label{fig:prodirect}
\end{figure}

\mypara{Our Vision}
We aim to bridge the gap between programmatic and direct
manipulation (DM) with hybrid, \emph{prodirect manipulation} systems
that support three interrelated modes of uses:
\textbf{(A)}
The output of a program should be directly manipulable by the
user while the system infers updates to the program, in real-time,
that matches the user's changes to the output. We refer
to this as \emph{live synchronization};
\textbf{(B)}
Program fragments should be inferred, or
\emph{synthesized}, automatically from desired output examples;
and
\textbf{(C)}
When the user wants to make a significant change to the output of a
program, the system should temporarily allow the program to become
out of sync with the output, after which it should synthesize program updates
that reconcile the changes and also reuse as much of the original program
as possible.
We refer to this mode of use as \emph{ad hoc synchronization},
because it describes a slower, more expensive process, as compared to
the immediate interactivity of live synchronization.
Together, these three integrated modes of use
(depicted in \autoref{fig:prodirect})
will form a kind of
\emph{bidirectional programming}~\cite{DagstuhlBX}, in the sense that
either the program or its output can be manipulated
to achieve a desired result.

Two, usually distinct
``masses'' of users stand to benefit:
``programmers'' would enjoy more interactive,
assisted programming environments, and
``end users'' could customize systems in task-specific ways.
We believe that merging the two kinds of software systems
is particularly important given the increasing
prevalence of computational fluency among a broad
range of users.

\section{Motivating Example} \noindent
Imagine the task of creating a series of
Scalable Vector Graphics (SVG) designs to illustrate
the mechanics of ferris wheels.

\begin{figure}[t]
\centering
\includegraphics[scale=0.40]{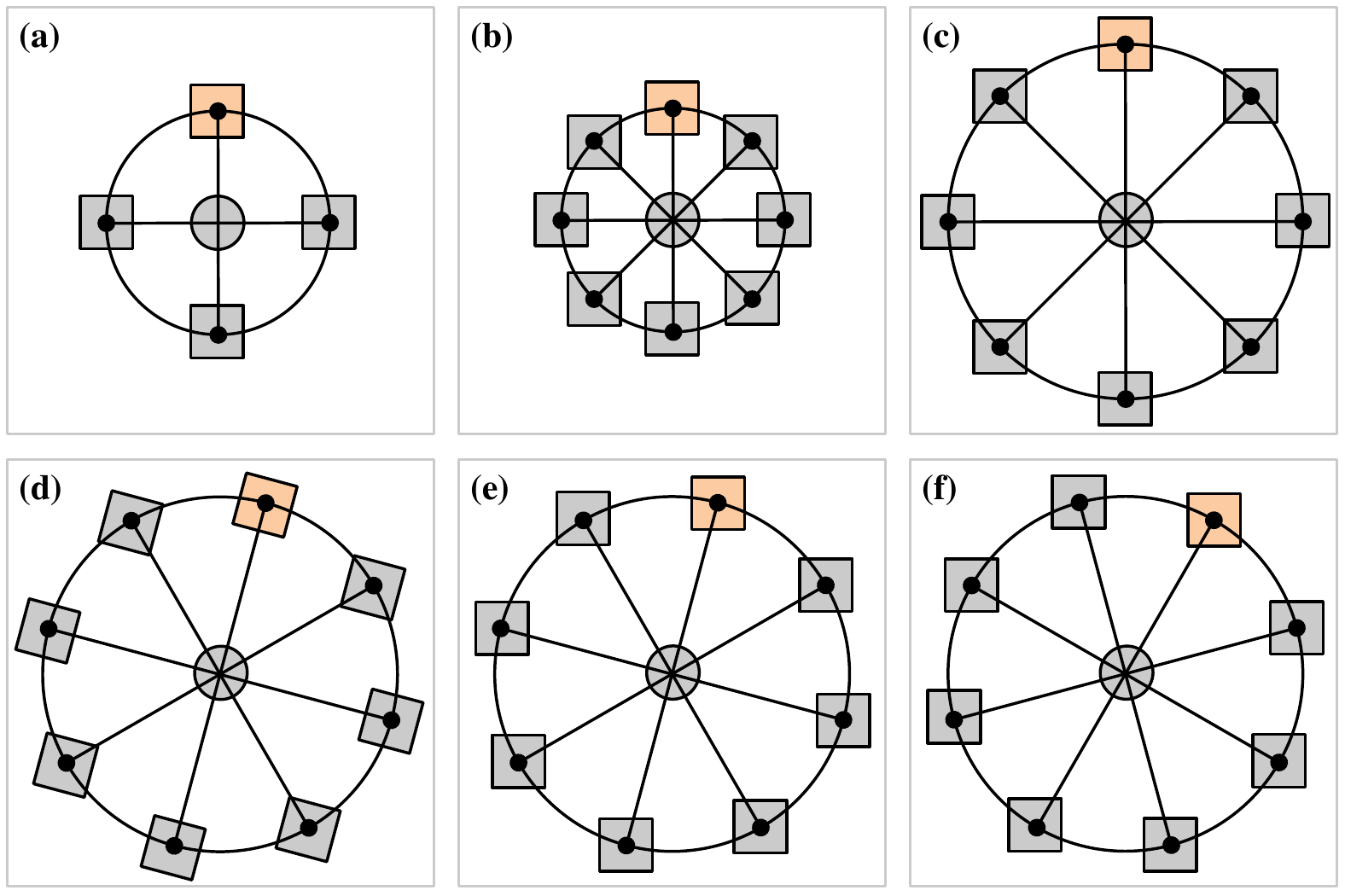}
\caption{Ferris Wheel Diagrams}
\label{fig:ferris}
\end{figure}

\mypara{Direct Manipulation (2015)}
Made in Microsoft \powerpoint{},
Diagram (a) of \autoref{fig:ferris}
shows a four-spoke wheel, constructed with several copied-and-pasted
elements that are manually placed by clicking-and-dragging and using
built-in snap-to-ruler operations. Similar kinds of GUI-based
edits lead to the eight-spoke version in Diagram (b).
Unfortunately, very little from the initial
design can be reused when adapting the
design for some number of spokes that is not a multiple of four,
a different spoke length (as in Diagram (c)), or a different
size for the passenger cars. Matters are harder still when
trying to depict the ferris wheel in motion. Simply rotating the
entire design (Diagram (d)) is unacceptable, as the passenger
cars ought to remain vertical throughout. Correctly rotated versions
(Diagrams (e) and (f)) require substantial
reimplementation of the initial design.

\mypara{Prodirect Manipulation (2015--2025)}
Instead, imagine a system that allowed
the user to carry out the same task as follows.
First, the user performs typical DM operations to create and style a
single square, copies and pastes it multiple times, and drags
the four squares approximately into a circle.
Next, the user asks the system to infer a small set of
candidate programs (say, fewer than three) that produce a result
similar to the user's drawing (Goal B) and chooses one of them.
Then, the user changes a single parameter in the program to draw ten
evenly spaced cars instead of four.
After the new output is rendered,
the user directly manipulates the size of one of the squares,
and the system immediately infers an update to the program such that
the sizes of all the squares change (Goal A).
After that, the user performs typical DM operations to draw lines
between a few pairs of squares, and then asks the system to
infer a new program fragment to generalize the new shapes
within the context of existing ones (Goal C).
Given a few options, the user chooses one where the number of
spokes and their rotated angles match that of the previously synthesized
circle of squares.
Equipped with this program, the user can directly manipulate visual
aspects of the wheel (\eg{} the size of its cars and spokes)
and programmatically manipulate aspects that are more difficult
to interpret visually (\eg{} the number of spokes).
Furthermore, the resulting high-level program can be reused
in other settings where it may be useful.

\section{Proposed Research} \noindent
Our goal is to develop new program synthesis algorithms
and user interface designs that achieve this kind of workflow
for a variety of domains with visual as well as textual data.

\mypara{Challenges}
Each of the three modes of use (\autoref{fig:prodirect}) poses technical
challenges that can be viewed
as a form of \emph{program synthesis}, which
is the task of inferring a program expression
based on some specification ---
a logical specification (\eg{}~\cite{Manna1980,SrivastavaPOPL2010,KuncakPLDI2010}),
positive and negative input-output examples~(\eg~\cite{KuncakCAV2011,GulwaniPLDI2011,GulwaniPLDI2015}), or
a partial program implementation called a
\emph{sketch}~(\eg{}~\cite{SketchingThesis}).
The two primary challenges for any program synthesis approach are
(i) efficiently exploring the large space of programs and
(ii) disambiguating among multiple programs that satisfy the specification.

These challenges are amplified for DM systems,
which must be responsive with very low latencies.
One common approach in \emph{programming by example} systems
(\eg{}~\cite{GulwaniPLDI2011,GulwaniPLDI2011geo,SinghPLDI2013})
for taming the challenges is to limit the
search space to programs in a domain-specific language.
This will not directly help with our goals, however, because many users
already expect to work with high-level, general-purpose programs
(\eg{}, \LaTeX{}, Slideshow~\cite{SlideshowICFP04}, Processing).

\subsection*{Our Approach} \noindent
We have identified two main strategies for tackling
the challenges associated with our vision.

\mypara{Real-Time Program Synthesis}
Programs can often be factored into a high-level,
logical structure plus low-level details (\eg{} constants) that have
natural direct representations in the output.
Furthermore, a user's DM actions play a unique role compared to
many programming by example systems, namely, that
a specific change introduces a new input-output constraint that ought
to be weighted particularly heavily, more so than the existing input-output
constraints from the previous program-output relationship.

Building on these two insights,
we propose that (a) the run-time behavior of a program
be \emph{traced}, or \emph{logged}, with information that describes
how output values are computed, and that (b) the synthesis procedure
solve for ways that the same program execution, if followed again,
would produce the desired updated value, regardless of its effect
on values not manipulated by the user.
By limiting the space of
program updates in this way, we expect that it will be more likely
that desirable updates can be inferred quickly enough for a
responsive experience.

\mypara{Domain-Specific Synthesis of General-Purpose Programs}
We are interested in developing synthesis
techniques for general-purpose languages, in
contrast to domain-specific ones --- for example,
restricted languages of drawing commands for geometry
constructions~\cite{GulwaniPLDI2011geo},
regular expressions for spreadsheet transformations~\cite{GulwaniPLDI2011},
or low-level graphics primitives~\cite{QuickDraw2012}.
We base our goal on the notion that high-level programs
can be continually developed and extended by the user
as needs grow.
Our search procedures, however, will be domain-specific.
As a simple first step, one can
``lift'' the domain-specific programs from prior approaches into
general-purpose ones. More important, however, will be new
heuristics for generating high-level, readable
programs that can be further manipulated and extended by the user
depending on the setting.
For each application domain, we plan to study how users employ
both existing programmatic as well as direct manipulations systems
in order to design synthesis algorithms that produce useful programs
for a variety of common scenarios.

\subsection*{Applications} \noindent
We now discuss several example application domains.

\mypara{SVG}
As a first step towards our vision, we are developing the
\sns{} prodirect manipulation editor for SVG
(\url{ravichugh.github.io/sketch-n-sketch}).
In our work so far~\cite{prodirect},
we provide only the live synchronization mode of use
(Goal A from \autoref{fig:prodirect}),
which means that the user must start by writing a full program, after
which the output can be directly manipulated.
In \sns{}, the program and output
are displayed side-by-side; manipulating either half results
in an immediate update to the other, based on an initial
form of trace-based program synthesis.
Even without the remaining modes of use (Goals B and C), we have
used \sns{} to design and edit a variety of designs
(such as the ferris wheel)
that would be difficult using either programmatic or direct
manipulation.

\mypara{Data Visualization}
Many tools help researchers and data scientists create data
visualizations to work with various kinds of data sets.
As with the SVG domain, many of these solutions heavily
favor either programmatic or direct manipulation.
D3~\cite{D3} is a widely popular JavaScript library for
creating complex, interactive Web-based visualizations.
All of the desired interactivity in a particular output, however,
must be {explicitly} designed by the (expert) D3 programmer.
Instead, it would be preferable for certain properties of the output
to be interactive ``for free,'' as a result of live synchronization.
This could make it easier for expert programmers to focus on
implementing the more interesting features, and could provide
users with less programming expertise more control over the
output of their tools.

\mypara{Documents}
Authoring primarily textual documents is a ubiquitous task
carried out by a broad range of users. Similar in breadth is
the range of DM tools (\eg{} Microsoft Word),
programmatic tools (\eg{} \LaTeX{} and languages that
generate HTML+CSS), and some in between (\eg{} Adobe
InDesign and Google Web Designer).
Compared to the primarily-visual
domains discussed above, the subtleties of rendering text and other
layout information introduce more complex notions of mapping
a programmatic input into the final output --- for example,
spacing between characters and words in order to realize full
justification, and relative positioning to fit within current
window dimensions. These issues make the dichotomy between
programmatic and direct manipulation particularly acute;
even expert programmers often resort to an ad-hoc combination
of separate tools and manipulation of low-level formats,
such as HTML and CSS~\cite{Wang2012}.
These concerns are further exacerbated in the spreadsheet
document domain, where semi-structured data and programs
(\ie{} formulas) are thrown into the
mix~\cite{GulwaniPLDI2015}.

\mypara{Software Maintenance}
Consider the challenges of
maintaining collateral evolutions across program snippets that originated
from copy-and-pasted source code (\eg{}~\cite{CollateralEvolution});
repairing all related root causes of a bug (\eg{}~\cite{AngelicDebugging});
performing systematic edits (\eg{}~\cite{Sydit,LASE}); and
keeping multiple related configuration files consistent (\eg{}~\cite{Tang2015}).
In each of these domains, one can imagine high-level abstractions
(\ie{} programs that operate on other programs, or program transformations)
that capture the relationships in the low-level formats
(\ie{} individual programs). But ``directly manipulating'' the invidual
programs is often required or preferred, just like with
directly manipulating visual data. Viewed in this light, we
see opportunities for prodirect manipulation software engineering tools.

\subsection*{Implications} \noindent
We describe two factors that stem from our target application
domains and our proposed approach.

\mypara{Cross-Domain Interoperability}
Many applications involve data that cross boundaries between
multiple distinct software systems. For example, consider an SVG image
that is designed in isolation, then included in a slideshow
where it is cropped to better fit the context, which is later compiled
to PDF. Ideally, we could track the provenance of the SVG image throughout
all of these intermediate steps, so that changes to the original SVG image
propagate all the way to the final PDF, and vice versa.
Assuming prodirect manipulation systems for each of the constituent
phases, having general-purpose programs produce output values with
lightweight run-time traces may provide a suitable
``common exchange format'' for reusing high-level abstractions
across traditionally disparate systems.

\mypara{Evaluation Criteria}
Many synthesis problems come with clear notions of
correctness, such as logical specifications or input-output
examples (\ie{} test cases) that a program ought to satisfy.
Furthermore, when restricted to a domain-specific language, 
search procedures often provide guarantees about completeness
(exhausting the search space) and quality of the solution
(finding the best candidate).

Many of our synthesis problems will have neither property.
First, in many domains (particularly visual ones), there are few
clear notions for precise specification.
By analogy, we can understand
that, today, working directly with low-level formats such as
SVG, HTML, and CSS constitutes working with the ``assembly languages''
of these domains. By developing prodirect manipulation
systems that enable and promote working at higher levels of
abstraction, it is likely for common patterns to emerge that
will form the basis of future specifications.
Second, because we aim to synthesize programs in a
general-purpose language, we have little hope for providing strong
optimality guarantees.

In light of these challenges, as well as our ultimate goal to provide
novel user interfaces to a broad range of users
(programmers and end users), we intend to base our evaluations
primarily along more qualitative measures. This will include detailed
case studies and user studies, to shed light on how the new systems
compare to previous ones. Despite differences among application domains,
we expect to develop insight for how to explain synthesized program
fragments and their behaviors to users in intuitive ways,
which may be reused across program synthesis domains.

\section{Related Work} \noindent
To conclude, we briefly mention several additional projects
that are related to our main vision.

\mypara{GUIs that Generate Code}
Some DM tools generate ``code behind'' GUI-based representations
(\eg{} Garnet~\cite{Garnet} for user interfaces and
QuickDraw~\cite{QuickDraw2012} for beautifying
vector graphics), but the generated artifacts are typically just as
low-level as the target format itself. In contrast, we aim to generate
code in high-level programming languages.

\mypara{Constraint-Oriented Programming}
Constraint-oriented systems, such as
SketchPad~\cite{SketchpadThesis} and ThingLab~\cite{Borning1981},
include declarative programming models where constraint
solvers are used during evaluation.
Although the interactivity provided by such systems
is similar to the live synchronization we propose,
we aim to work in the context of more traditional, deterministic
programming models, which are more widely found in practice
than constraint-oriented ones.

\mypara{Live Programming}
The general goal of live programming systems (\eg{}~\cite{McDirmid2013})
is to provide immediate feedback about changes made to the
program. We share this goal, in addition to supporting the reverse
direction.

\mypara{Bidirectional Programming}
Bidirectional programming languages (see~\cite{DagstuhlBX} for an overview
survey) allow users to transform data in either of two directions, often
using domain-specific language constructors or domain-specific
knowledge in reconciling ambiguities.
Our high-level goal is similar in spirit,
except that we aim to use more general-purpose
programming languages in specific application domains.
Furthermore, unlike many language-based approaches~\cite{DagstuhlBX},
there may not necessarily be clear notions
of correctness for many domains, so more subjective notions of
usability and utility will be particularly important.

\section*{Acknowledgments}
\noindent
The author wishes to thank
Jacob Albers, Brian Hempel, and Mitch Spradlin
for collaborating on \sns{},
and Shan Lu for suggesting improvements to this paper.

\end{document}